\documentclass[11pt,njp]{iopart}
\usepackage{iopams}
\expandafter\let\csname equation*\endcsname\relax 
\expandafter\let\csname endequation*\endcsname\relax 
\usepackage{amsmath}

\newcommand{\nc}{\newcommand}
\nc{\eq}{\begin{equation}}
\nc{\eeq}{\end{equation}}
\nc{\eqa}{\begin{eqnarray}}
\nc{\eeqa}{\end{eqnarray}}
\nc{\nn}{\nonumber}
\def\bra#1{\mathinner{\langle{#1}|}}
\def\ket#1{\mathinner{|{#1}\rangle}}

\usepackage{graphicx,subfigure}
\usepackage{times}

\usepackage{graphicx}
\usepackage{color}
\usepackage[usenames,dvipsnames]{xcolor}

\begin{document}

\title{\sf \bfseries Dynamical decoupling efficiency versus quantum non-Markovianity}

\author{\sf \bfseries Carole Addis$^{1}$, Francesco Ciccarello$^{2}$, Michele Cascio$^{3}$, G. Massimo Palma$^{2}$, Sabrina Maniscalco$^{4}$}

\address{$^{1}$SUPA, EPS/Physics, Heriot-Watt University, Edinburgh, EH14 4AS, UK\\
$^{2}$NEST-INFM (CNR) and Dipartimento di Fisica e Chimica, Universit\`a degli Studi di Palermo, Via Archirafi 36, I-90123 Palermo, Italy\\
$^{3}$Dipartimento di Fisica e Chimica, Universit\`a degli Studi di Palermo, Via Archirafi 36, I-90123 Palermo, Italy\\
$^{4}$Turku Center for Quantum Physics, Department of Physics and Astronomy, University of Turku, FIN-20014 Turku, Finland}
\ead{ca99@hw.ac.uk}
\begin{abstract}
We investigate the relationship between non-Markovianity and the effectiveness of a dynamical decoupling protocol for qubits undergoing pure dephasing. We consider an exact model in which dephasing arises due to a bosonic environment with a spectral density of the Ohmic class. This is parametrized by an Ohmicity parameter by changing which we can model both Markovian and non-Markovian environments. Interestingly, we find that engineering a non-Markovian environment is detrimental to the efficiency of the dynamical decoupling scheme, leading to a worse coherence preservation. We show that each dynamical decoupling pulse reverses the flow of quantum information and, on this basis, we investigate the connection between dynamical decoupling efficiency and the reservoir spectral density. Finally, in the spirit of reservoir engineering, we investigate the optimum system-reservoir parameters for achieving maximum stationary coherences.

\end{abstract}

\section{Introduction}
Dynamical decoupling (DD) techniques for open quantum systems are among the most successful methods to suppress decoherence in qubit systems \cite{ref1Lorenza, ref2Lorenza}. Sophisticated control design have superseded earlier schemes such as the so-called  ``bang bang" periodic dynamical decoupling (PDD) \cite{ref2Lorenza} and its time-symmetrized version \cite{Lorenza1}-\cite{Lorenza2a}. On the one hand concatenated DD schemes (CDD) have been developed to counter decoherence for general noise scenarios \cite{CDDs, CDDsa}, on the other hand optimal approaches to minimize errors in specific noise settings have been discovered \cite{CDD}. In both cases, a high sensitivity of the efficiency of the protocols to the pulse timing has been demonstrated. As the performance of all DD schemes crucially depends on the timescale of the environment correlation function, it is clear the important role played by spectral properties of the noise causing decoherence and introducing errors \cite{Daniel}. In \cite{Paper}, an exactly solvable pure dephasing model was used to compare the efficiency of certain DD protocols in Ohmic, sub-Ohmic and super-Ohmic environments. This is important because of the increasing ability to engineer experimentally environmental properties, such as the spectral distribution \cite{eng}. 

During the last few years, a new perspective into the meaning and importance of non-Markovian dynamics has emerged. This has led to intense activity on fundamentals of open quantum systems, and non-Markovian systems in particular, culminating with the introduction of new definitions and characterisations of non-Markovianity. These definitions and their implications are reviewed in \cite{Rev1} and \cite{Rev}. In this new framework, non-Markovian is no longer simply synonymous  with persistence of system-environment correlations with the latter turning out to be a necessary but not sufficient requirement for non-Markovianity.

The advantage of such new approaches to non-Markovianity is to allow for a quantitative assessment of the usefulness of memory effects in quantum technologies, as they closely follow the formalism of information theory.
A number of results, indeed, support the idea that non-Markovian dynamics is most suitable for quantum communication and information processing purposes \cite{P1}-\cite{P6}. Moreover, very recently it has been investigated in \cite{Susanna} how non-Markovianity affects the effectiveness of optimal-control strategies in the case of amplitude-damping-type channels, finding the existence of regimes where non-Markovianity can be either beneficial or detrimental. 

In the present paper we re-examine a simple example of a dynamical decoupling scheme for a decohering channel in the light of the above mentioned new approach to non-Markovianity. In particular we will assess the performance of DD in the presence of information back-flow, a common quantifier of non-Markovianity.
Futhermore, in analogy with the perspective which views DD as a way to engineer environment spectra, here  we also study if and how the DD pulses change the non-Markovian character of the dynamics, e.g. whether they induce information back-flow, as defined in \cite{BLP}.

The structure of the paper is the following. In section 2 we review the basic definitions of non-Markovianity recently adopted by the open quantum systems community, and we motivate this approach. In section 3 we introduce the system of interest, namely the pure dephasing model including its exact solution in presence of periodic dynamical decoupling. In section 4, we discuss how the DD pulses affect information flow and hence modify the Markovian/non-Markovian character of the dynamics. In section 5, we investigate whether non-Markovian or Markovian dynamics are best suited to DD, i.e., lead to optimal performance. In section 6, we discuss in the spirit of reservoir engineering, the optimum system parameters for achieving maximum stationary coherences. Finally in section 7 we summarize our findings and draw the conclusions.

\section{Non-Markovian quantum dynamics}
The Born-Markov approximation is a cornerstone in the treatment of the dynamics of open quantum systems.  Under the assumption of a weak system-environment coupling one can safely assume that the environment is hardly modified by its coupling with the system. Furthermore ``big" environments are characterized by very short self correlation times. In this scenario the environment does not keep memory of the state of the system. The master equations so obtained describe the coarse grained system time evolution, i.e. the system dynamics on timescales larger than the environment correlation time. In this framework one refers to an open quantum system dynamics as Markovian when one neglects the correlations that build up between system and environment. However in several circumstances, the timescales over which the environment keeps memory of the system are finite.  A paradigmatic example of open system dynamics where one can study exactly the build up of correlations between system and environment is pure decoherence \cite{puremodel1}-\cite{puremodel3}. Such system is indeed analytically soluble and one can show how the timescale of such build up is related to the environment spectral density. In this framework, non-Markovianity is a property of the noise acting on the system and its signature is the  persistence of system-environment correlations, typically associated to structured spectral density of the reservoir \cite{NMDD1}-\cite{overlap}. Indeed the partial persistence of such correlations is the key ingredient of DD

In the quantum information theory approach to open quantum systems, however, non-Markovianity is a property which characterizes the time evolution of the open quantum system, more precisely its {\it dynamical map}. $\Phi_t$. By definition, if the open system is initially in a state $\rho_0$, its state at a later time $t$ is given by $\rho(t)\!=\!\Phi_t\rho_0$. Hence, $\Phi_t$ embodies a $t$-parametrized family of quantum channels. In some special cases this property can be also related to the form of the generalized master equation. The dynamical map, and hence the open system dynamics, does not depend, however, only on properties of the environment, but crucially also on the type and strength of interaction between system and environment.
Therefore, one cannot properly talk of non-Markovian environments because the system-environment interaction Hamiltonian also plays a key role. A second point worth recalling is that, as non-Markovianity is a property of the dynamical map, it cannot depend on the initial state considered.

With these consideration in mind we will now briefly recall the main motivations which have led to the new definitions of non-Markovianity. The trace-distance measure of \cite{BLP}, which paved the way to all the others, stems from the following desiderata: (i) to give a physical interpretation of memory effects in terms of information back-flow; (ii) to define non-Markovian dynamics independently from the specific structure of the master equation of the system. The underlying idea was to have a definition that was not based on mathematical properties but rather on the occurrence of physical effects, such as revivals of the information content of a quantum open system. This should be contrasted with the more mathematical approach according to which Markovian dynamics is described by divisible dynamical maps, i.e., maps satisfying the property $\Phi_t=\Phi_{t,s} \Phi_s$, with $\Phi_{t,s}$ completely positive and trace preserving (CPTP). Non-Markovian dynamics occurs instead when $\Phi_{t,s}$ loses complete positivity \cite{RHP, Rev1}. 

The intuitive notion of information back-flow in Markovian and  non-Markovian systems and of its reversal by means of DD is well established (see e.g. \cite{EP, EPa}). However only recently the concept of information flow in open system dynamics has been rigorously quantified using information theoretical quantities such as, e.g., distinguishability between states \cite{BLP}, coherent information \cite{P6}, Fisher information \cite{FI}, mutual information between input and output state of a channel \cite{MI}, fidelity \cite{F}, and so on and so forth. Correspondingly a number of non-Markovianity measures or witnesses have been proposed based on the temporal behaviour of these quantities. 
The key property exploited in these definitions is that the time evolution of any of these quantities, say distinguishability between quantum states, is contractive under CPTP maps. Hence a temporary increase of distinguishability, which is physically interpreted as a partial increase in the information content of the open system due to memory effects, always implies that divisibility of the dynamical map is violated.

Let us indicate with $I_{\rho_0} (t)$ a quantifier of the information content of the system. This generally depends on the initial state $\rho_0$ (or in some cases on pairs of initial states) and, due to contractivity, it is such that $I_{\rho_0} (t) \le I_{\rho_0} (s)$, for any $s \le t$. The non-Markovianity measure ${\cal N}_{I}$ is now defined as
\eq
{\cal N}_{I} = \max_{\rho_0}\int_{\sigma} \frac{d I_{\rho_0} (t')}{dt'}  dt',\label{NM-meas}
\eeq
where the integral is defined over all time intervals for which $ d I_{\rho_0} (t)/dt >0$. The quantity $d I_{\rho_0} (t)/dt $ defines information flow. Hence, information back-flow is indicated by positive values of the derivative of $I_{\rho_0} (t)$. 

A remarkable consequence of defining non-Markovianity on this refined basis compared to its traditional notion is that Eq. (\ref{NM-meas}) predicts that some known time-non-local master equations are indeed Markovian, i.e., they entail a dynamical map $\Phi_t$ such that ${\cal N}_{I}\!=\!0$ \cite{tnl}.

The use of this new characterisation of non-Markovianity has allowed one to prove that memory effects, defined as revivals of information theoretical quantities, are useful for quantum technologies \cite{QKD, P6, QT1, QT2}, they directly control the lower bound of uncertainty relations \cite{Goktag}, and have a powerful thermodynamical meaning in terms of revivals of extractable work \cite{Bognaarx}. Due to these reasons we believe that the connection between non-Markovianity as defined in this section and dynamical decoupling is worth exploring. This will be the topic of the following sections.

It is worth mentioning that, while the variety of non-Markovianity measures mentioned above in general do not coincide, for a qubit undergoing {\it pure dephasing} (i.e., the dynamics studied in this paper) they all consistently witness non-Markovian behaviour \cite{BLP4}.

\section{The System}
Let us consider the following microscopic Hamiltonian describing the local interaction of a qubit (i.e., a two-level system) with a bosonic reservoir, in units of $\hbar$ \cite{puremodel2, bp},
\eq
H=\omega_0\sigma_z+\sum_k\omega_k a^\dagger_k a_k+\sum_k\sigma_z(g_ka_k+g_k^*a_k^{\dagger}),
\eeq
with $\omega_0$ the qubit frequency, $\sigma_z$ the usual $z$-component of the qubit pseudospin, $\omega_k$ the frequency of the $k$th reservoir mode, $a_k (a_k^\dagger)$ the corresponding annihilation (creation) operator and $g_k$ the coupling constant associated with the qubit-$k$th mode interaction. 
This model can be solved exactly \cite{puremodel1}-\cite{puremodel3}. For factorised initial conditions and in the interaction picture, the master equation for the qubit density matrix $\rho$ is given by
\eq
\dot\rho=\gamma_0(t)[\sigma_z\rho\sigma_z-\rho]/2,
\eeq
the solution of which yields decay of the coherences (pure dephasing) as follows
\eq
\rho_{01}(t)=\rho_{10}^*(t)=\rho_{01}(0)e^{-\Gamma_0(t)},\label{rho01}
\eeq
where 
\eq
\gamma_0(t)=d \Gamma_0(t)/dt, \label{gamma}
\eeq and
\eq
\Gamma_0(t)=\int_0^{\infty}\frac{I(\omega)}{\omega^2}[1-\cos(\omega t)]d\omega. \label{eq:Gamma0}
\eeq
Here, $I(\omega)=\sum_j\delta(\omega-\omega_j)|g_j|^2$ is the spectral density function characterizing the interaction of the qubit with the oscillator bath (this is assumed to be at zero temperature). We consider the widely studied class of spectral densities of the form \cite{fran}:
\eq
I(\omega)=\alpha\frac{\omega^s}{\omega_c^{s-1}}e^{-\omega/\omega_c},\label{SDohm}
\eeq
with $s$ the Ohmicity parameter, $\alpha$ a dimensionless coupling constant and $\omega_c$ a cutoff frequency. Ohmic spectrum corresponds to $s=1$, while super-Ohmic spectra correspond to $s>1$ and sub-Ohmic to $s<1$. 
The expression for $\Gamma_0(t)$ can be calculated analytically by inserting Eq.~(\ref{SDohm}) into Eq.~(\ref{eq:Gamma0}) and is given, for super- and sub- Ohmic spectra, by \cite{HuelgaMet,Greg}
\eqa
\Gamma_0(t)&=&\frac{\alpha\tilde \Gamma[s]}{s-1}\left[ 1-(1+t^2)^{-s/2}\cos(s\arctan(t))+t\sin(s\arctan(t)) \right],
\label{eq}
\eeqa
with $\tilde \Gamma[s]$ the Euler Gamma function. We note that Eq. (\ref{eq}) is written in dimensionless units by introducing $\omega_c^{-1}$ as a time scale. For an Ohmic spectrum, we can write Eq.~(6) as,
\eq
\Gamma_0(t)=\frac{\alpha}{2}\ln(1+t^2).
\eeq

Let us now address the qubit behaviour in the presence of an arbitrary sequence of instantaneous bang-bang pulses, each of which being modelled as an instantaneous $\pi$-rotation.
In such a case, the decoherence process of the qubit can still be exactly described by replacing $\Gamma_0(t)$ in Eq.~(\ref{rho01}) with a modified decoherence function $\Gamma(t)$ \cite{Paper}.  An exact representation of the controlled decoherence function in terms of its free (uncontrolled) counterpart has been obtained in \cite{Paper}. 
Consider an arbitrary storage time, $t$, during which a total number of $N$ pulses are applied at instants $\{t_1,...t_n,...t_f\}$, with $0<t_1<t_2<...<t_f<t$. As shown by Uhrig \cite{U, Ua}, the controlled coherence function $\Gamma(t)$ can be worked out as,
\eq
\Gamma(t)= \left\{ 
  \begin{array}{l l}
    \Gamma_0(t) & t\leq t_1 \\
    \Gamma_n(t) & t_n<t\leq t_{n+1},0<n<N \\
    \Gamma_N(t) & t_f<t \label{eq:Gamma} 
  \end{array} \right.,
 \eeq
where, for $1\leq n \leq N$, 
\eqa
\Gamma_n(t)&=&2\sum_{m=1}^n(-1)^{m+1}\Gamma_0(t_m)\nn\\&+&4\sum_{m=2}^n\sum_{j<m}\Gamma_0(t_m-t_j)(-1)^{m-1+j}\nn\\&+&2\sum_{m=1}^n(-1)^{m+n}\Gamma_0(t-t_m)+(-1)^n\Gamma_0(t).
\label{Gamma}
\eeqa

In the next section, we will use $\Gamma(t)$ as given above to investigate the dynamics in terms of information flow and quantum non-Markovianity. In the following we set the dimensionless coupling constant $\alpha$, appearing in Eqs. (7)-(9), to unity as the introduction of this front-factor only leads to a rescaling of both the decoherence factor $\Gamma_0(t)$ and the controlled function $\Gamma(t)$. Since we are interested in comparing decoherence in different scenarios (that is, e.g., unpulsed case with DD-pulsed case), this factor is irrelevant. Although this is true for pure dephasing, the coupling constant plays a role for other open systems dynamics, e.g. dissipative dynamics,  where a perturbative analysis of the decay is needed. We note that while this expression for $\Gamma(t)$ has been derived for a qubit interacting with a quantum bosonic bath, its exact representation holds also for arbitrary Gaussian phase randomisation processes. Hence the main conclusions of our study are applicable also to experimental settings such as trapped ions \cite{exp2},\cite{exp2a} and solid-state qubits \cite{NMDD3}, \cite{exp2b},\cite{exp2c}.

\section{Pulse-induced information flow reversal} 

To characterize the open dynamics under study from the viewpoint of information flow, we make use of a well-known measure of non-Markovianity introduced in \cite{BLP} and known as the BLP or trace-distance measure. This is based on the time evolution of the trace distance between a pair of initial states of the open system, this being a measure of their relative distinguishability. In a Markovian process the distinguishability between any two quantum states decreases monotonically in time, indicating a loss of information about the system due to continuous monitoring of the environment. In a non-Markovian process, in contrast, it can partly regrow for some time intervals, indicating information back-flow into the system. For the system here considered the non-Markovianity measure of \cite{BLP} has a simple analytical expression \cite{anaone}: 
\eq
\mathcal{N}=-\int_{\gamma<0}dt\:\gamma(t)e^{-\Gamma(t)}, \label{eq:N}
\eeq
where $\gamma(t)$ [cf.~Eq.~(\ref{gamma})] is the modified decoherence rate and the integral, as suggested by our notation, is extended over the time intervals such that $\gamma(t)\!<\!0$.
Hence, one can immediately associate information back-flow with negative values of $\gamma(t)$.  The non-Markovianity defined in this way for the free system has been studied previously (see \cite{BLP1}). In particular, it was shown analytically that, for $\Gamma(t)\!=\!\Gamma_0(t)$ and in the case of spectral density (\ref{SDohm}), the measure takes non-zero values if and only if $s>2$ \cite{Pinja}.
This  result establishes a connection between the definitions of non-Markovianity discussed in section 2 -- in particular the BLP measure -- and the form of the environmental spectral density for the Ohmic class given in Eq.~(\ref{SDohm}). Having this in mind, we sometimes refer in the following to Markovian (non-Markovian) dynamics to indicate the $s\le 2$ ($s>2$) regimes. 

The controlled decoherence function $\Gamma_n(t)$ [cf.~Eq.~(\ref{Gamma})]  may be rewritten as: 
\eq
\Gamma_n(t)=-\Gamma_{n-1}(t)+2\Gamma_{n-1}(t_n)+2\Gamma_0(t-t_n).
\eeq
Hence, it is straightforward to define a relation connecting $\gamma_n(t_n)$, namely $d\Gamma_n/dt$ at the moment $t_n$ when the system is pulsed, and the corresponding quantity at the previous instant \cite{Paper}: 
\eq
\gamma_n(t_n)= -\gamma_{n-1}(t_n),\label{gammannm1}
\eeq
where for $1\leq n \leq N$, 
\eq
\gamma_n(t)=2\sum_{m=1}^n (-1)^{m+n}\gamma_0(t-t_m)+(-1)^n\gamma_0(t).
\eeq
Interestingly, Eq.(\ref{gammannm1}) clearly shows that information flow is reversed whenever a pulse interrupting the free-system dynamics is applied. Hence, strictly speaking, a Markovian open system dynamics will always be turned by the pulsing into a non-Markovian one (although in some cases the resulting non-Markovianity measure can take negligible values).
\begin{figure}[h]
\begin{centering}
\includegraphics[width=0.88\textwidth]{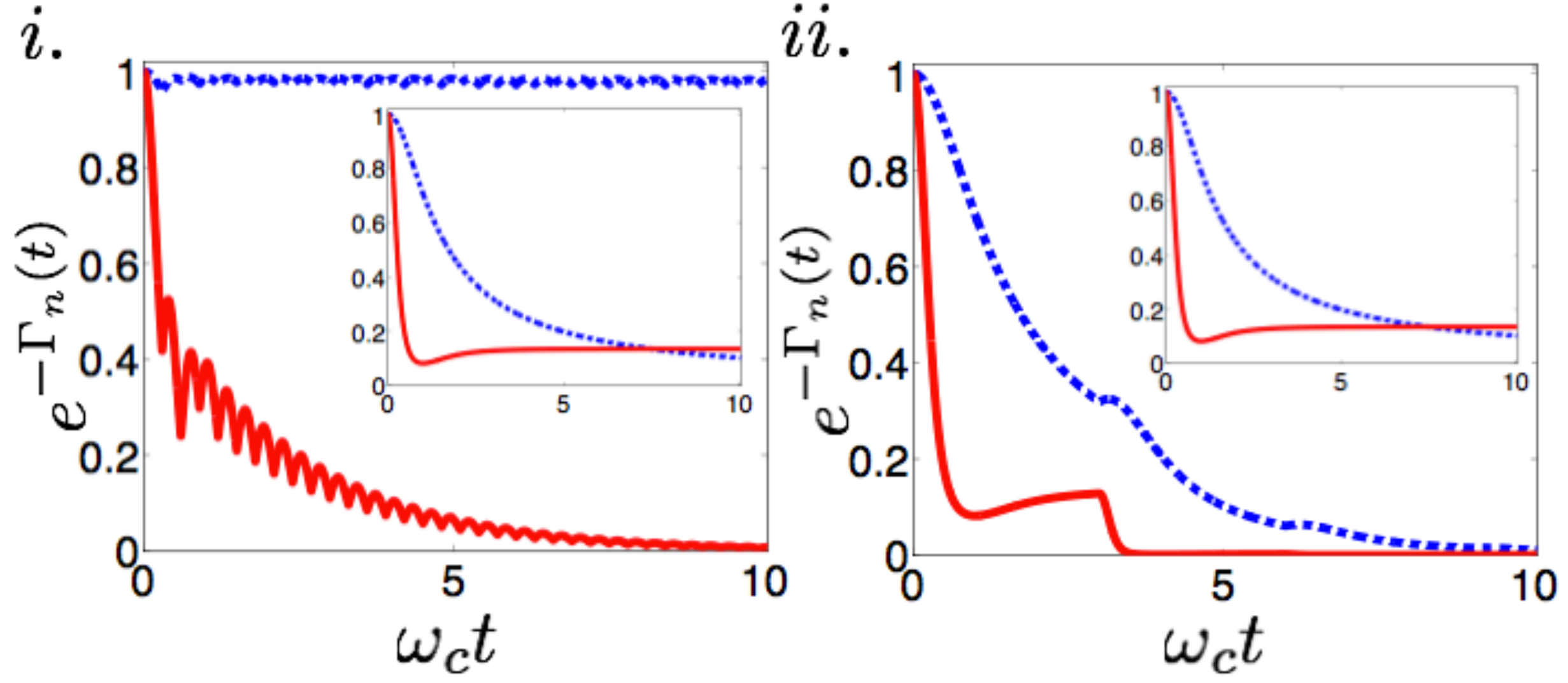}
\caption{(Color online) Time evolution of the controlled coherence $e^{-\Gamma_n(t)}$ for: i.~$\Delta t$=0.3 (short pulse spacing regime) and ii.~$\Delta t$=3 (long pulse spacing regime), in units of $\omega_c^{-1}$. The Ohmicity parameters are $s=1$ (blue dashed line) and $s=4$ (red solid line), which are respectively an instance of Markovian and non-Markovian dynamics. In the inset, we display the free uncontrolled coherences $e^{-\Gamma_0(t)}$, which shows in particular that $\bar{t}\!=\!1$ for $s\!=\!4$. All times are expressed in units of $\omega_c^{-1}$.}
\label{figure1}
\end{centering}
\end{figure} 

In figure 1, we study the time evolution of coherences of the purely dephasing system subject to periodic DD. 
For the sake of simplicity, we focus here on equally-spaced DD pulses applied at times $t_n=n \Delta t $, with $n=1,2,3,...$. We call $\bar{t}$ the first time instant at which $\gamma_0(\bar{t})=0$, i.e., after which information flow is temporarily reversed (in the unperturbed dynamics). This time always exists for $s>2$ \cite{Pinja}. For divisible (Markovian) dynamics, in the Zeno regime, the shorter is the interval between the DD pulses the higher is the efficiency of the DD scheme \cite{Zeno2, Zeno3}. For non-Markovian ones the same holds, provided that $\Delta t<\bar t$. This is a straightforward consequence of the quantum Zeno effect whose connection with dynamical decoupling has been shown in \cite{Zeno1}. In figure 1, we consider the cases $s\!=\!1$ and $s\!=\!4$ as paradigmatic instances of a Markovian and non-Markovian (free) dynamics respectively. For each of them, we consider both the case of a short and large pulsing period (short-pulsing and long-pulsing regimes).

We first note that, for any time $t < \bar{t}$, the unperturbed coherences are always higher for $s \le 2$ (Markovian case) than for $s>2$ (non-Markovian case). Since the effect of the pulses is always to reverse information flow and therefore preserve better coherences, we conjecture that, in the short-pulsing regime, Markovian Ohmic environments are more favourable to protect coherences via DD compared to non-Markovian ones. 
In the Markovian case, however, DD inhibits loss of coherence compared to the unpulsed free evolution. 
This is confirmed by figure 1 i showing that the pulsing is fully successful in inhibiting the coherences decay for the $s\!=\!1$ case, while it is not in the $s\!=\!4$ case.

Instead, in the long-pulsing regime (see figure 1 ii), the efficiency of the DD scheme here considered is drastically reduced in both the Markovian and non-Markovian case and greatly depends on the details of the dynamics, hence no general conclusion can be drawn. In particular, in this regime, reversing information flow can have disastrous consequences for non-Markovian environments: If the first pulse occurs during a time of re-coherence (information back-flow), it will indeed induce an even faster coherences decay. This effect can be seen in figure 1 ii,  where in particular we study the case $s\!=\!4$ for a pulse spacing such that $\Delta t\!>\! \bar{t}$. One can note that the occurrence of the first pulse induces an extremely rapid deterioration of coherences (when compared to the unpulsed free dynamics).

\section{Efficiency versus Non-Markovianity measure}
To elucidate the relationship between the non-Markovian character of the free dynamics and the efficiency of dynamical decoupling techniques, we perfom a numerical investigation based on their respective measures. Most studies on DD quantify the efficiency of dynamical decoupling sequences by means of the fidelity function, measuring the overlap between the state at time $t$ and the initial state $\rho(0)=\ket{\Phi(0)}\bra{\Phi(0)}$, namely
\eq
\mathcal{F}(t)=\bra{\Phi(0)}\rho(t)\ket{\Phi(0)},
\eeq
where $1/2\leq \mathcal{F}\leq 1$. In the weak-coupling approximation, the coherences decay as $\mathcal{C}(t)=e^{-R(t)t}$ where $R(t)$ is the overlap interval of the noise spectral density and the filter function generated by the DD sequence \cite{overlap}. In this framework, fidelity is defined as:
\eq
\mathcal{F}(t)=\rho_{11}(0)^2+\rho_{22}(0)^2+2\rho_{12}(0)\rho_{21}(0)e^{-R(t)t}, \label{fidweak}
\eeq
with $\rho_{ij}$ the density matrix elements of the initial state. It is worth stressing that in this paper we use an exact approach that allows us to write the most general form of the decoherence factor as  $\mathcal{C}(t)=e^{-\Gamma(t)}$, with $\Gamma(t)$ given by Eq. (\ref{eq:Gamma}). In the weak coupling limit $\Gamma(t)$ reduces to $R(t)t$ and one obtains Eq. (\ref{fidweak}).

In the following, we are interested in quantifying how well the DD sequences protect the system from decoherence {\it at all times} and independently of the initial state. This is because we aim to study the efficiency of the DD scheme in connection to a property of the dynamical map, along the lines of what it is done when introducing non-Markovianity measures. Rather than the fidelity, which is both time-dependent and state dependent, we therefore introduce the following quantifier of DD efficiency:
\eq
\mathcal{D}(t_f)=\frac{\int_0^{t_f} e^{-\Gamma(t)}}{t_f} \label{eq:D}\,.
\eeq
The measure is bounded between zero (ineffective DD) and unity (ideal DD) and is based only on preserving the evolution of coherence undergoing dynamical decoupling up to some time $t_f$, which is assumed to be the duration of the DD pulsing scheme.
\begin{figure}[h]
\begin{centering}
\includegraphics[width=0.58\textwidth]{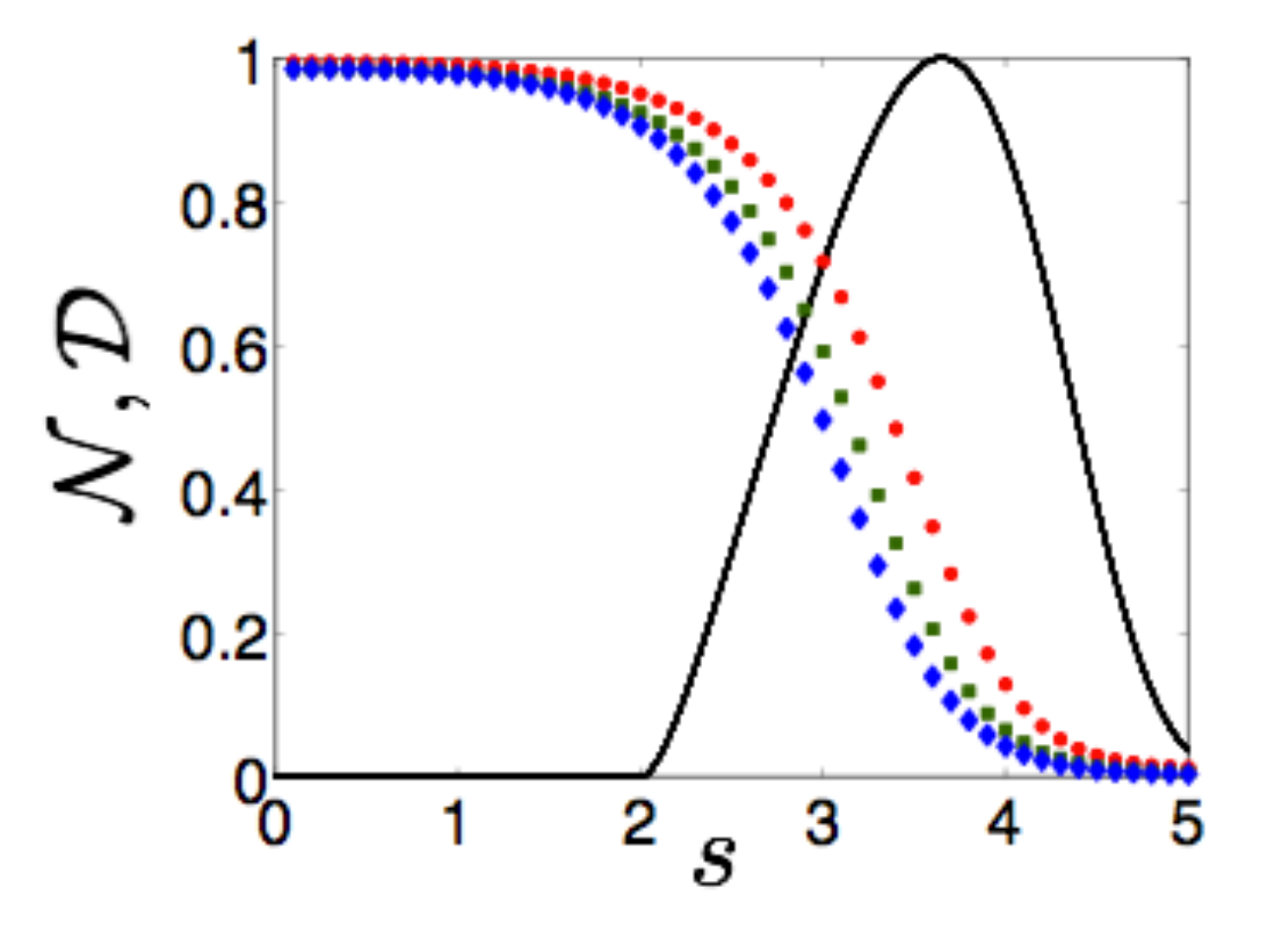}
\caption{(Color online) Non-Markovianity measure $\mathcal{N}$ of the free (unpulsed) dynamics (black solid line) and dynamical decoupling efficiency $\mathcal{D}$ against the Ohmicity parameter $s$ for $\Delta t=0.3$ (in units of $\omega_c^{-1}$). Dynamical decoupling efficiency is plotted for $\omega_c t_f=9.9$ (red dot), 19.8 (green square) and 30 (blue diamond). For comparison purposes, we have rescaled $\mathcal N$ to its maximum value.}
\label{figureD}
\end{centering}
\end{figure} 
In figure 2, we compare the DD efficiency measure $\mathcal{D}$, as defined by Eq.~(\ref{eq:D}) with $\Gamma (t)$ given by Eq.~(\ref{eq:Gamma}), and the non-Markovianity $\mathcal{N}$, as defined by Eq.~(\ref{eq:N}) with free decoherence $\Gamma_0 (t)$ given by Eq.~(\ref{eq:Gamma0}), as functions of the Ohmicity parameter $s$. We focus here on the short-pulsing regime where the efficiency of the DD scheme is the highest. The plot clearly shows a sharp decrease in $\mathcal{D}$ with the onset of non-Markovianity for $s > 2$. This quantity, however, is only sensitive to the Markovian to non-Markovian crossover ($s=2$) and not to the value of  $\mathcal{N}$ for $s>2$, as it keeps decreasing monotonically while  $\mathcal{N}$ has a clear peak around $s\!\simeq\!3.7$. For increasingly longer times $t_f$, the efficiency becomes increasingly sensitive to the onset of non-Markovian dynamics, indeed, for $t_f\rightarrow\infty$, we conjecture that $\mathcal{D}$ will decrease to smaller and smaller values for $s>2$. 

Note that the representative values of $t_f$ in figure 2 were chosen to be not too short. A too short value of $t_f$ will indeed yield an almost uniform behaviour of $\mathcal{D}$ as  function of the Ohmicity parameter $s$ since in such a case the coherences will still be high for any value of $s$. We have thus focused on the behaviour of $\mathcal{D}$ for sufficiently long times $t_f$, since this is actually what matter the most, namely, how to preserve coherences for long times. In figure 2, we have therefore considered increasingly long times until the limit of our computational capabilities. The plots clearly indicate a similar tendency.
\begin{figure}[h]
\begin{centering}
\includegraphics[width=0.6\textwidth]{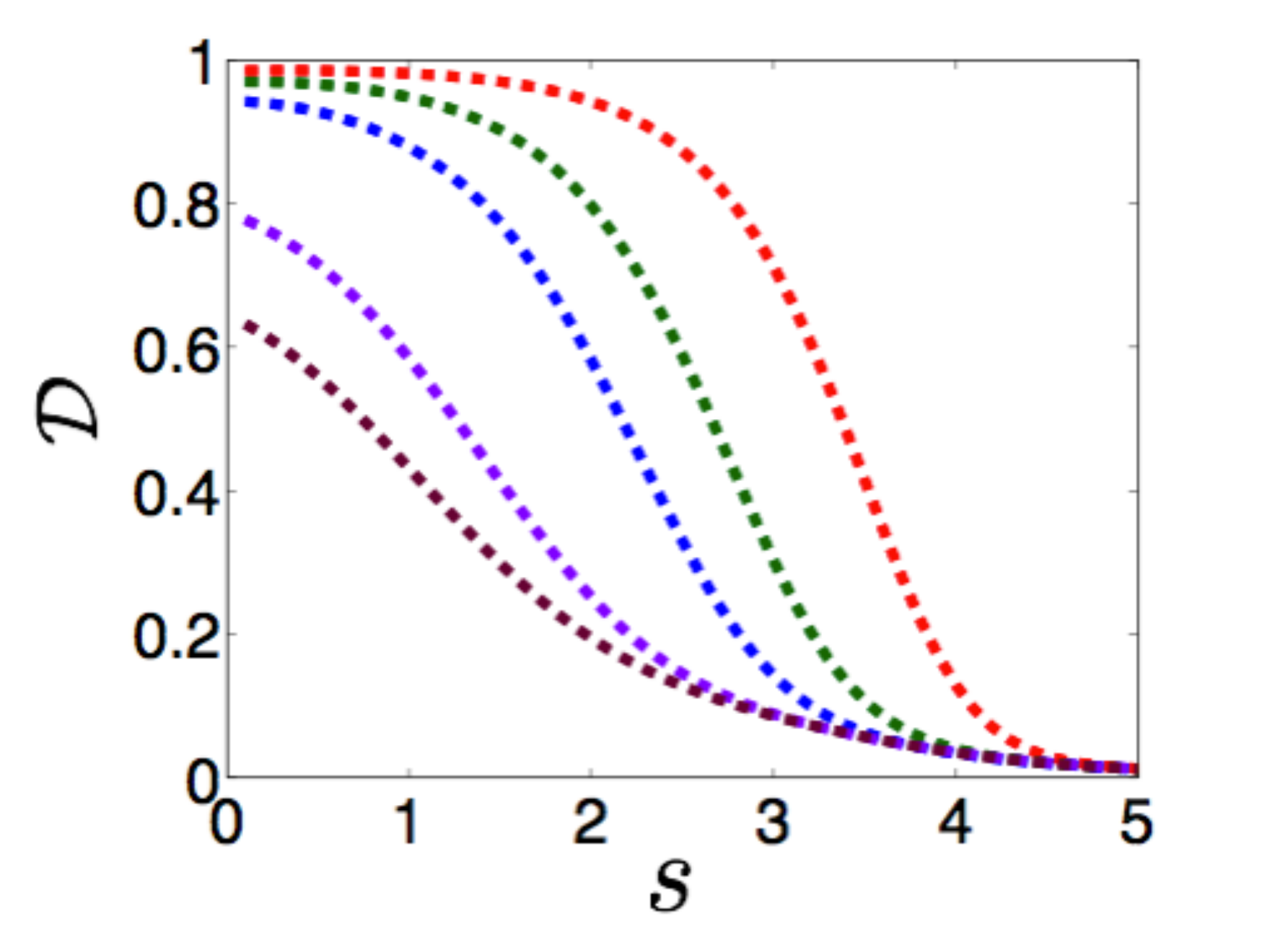}
\caption{(Color online) Dynamical decoupling efficiency $\mathcal{D}$ against the Ohmicity parameter $s$ for different values of the pulsing period (in units of $\omega_c^{-1}$): $\Delta t\!=\!$0.3 (red), $\Delta t\!=\!$0.4 (green) , $\Delta t\!=\!$0.5 (blue),  $\Delta t\!=\!$0.8 (purple) and $\Delta t\!=\!$1 (brown). The final pulse is applied at $t_f\!=\!N_{\text{max}}\Delta t\leq10~\omega_c^{-1}$ where $N_{\text{max}}$ is the maximum number of pulses that can be applied within the time interval $0\leq t\leq 10~\omega_c^{-1}$. }
\end{centering}
\end{figure} 
In figure 3, we study how the above comparison between the non-Markovianity measure and DD efficiency depends on the time intervals $\Delta t$. One can see that the behaviour displayed in figure 3  is rather insensitive to the time interval $\Delta t$. We have numerically checked that this conclusion is not dependent on the specific value chosen for $t_f$ in figure 3. Summarizing, figures 2 and 3 show that the maximum efficiency of PDD is obtained for pulse spacings $\Delta t < \bar{t}$ with Markovian dynamics ($s<2$). As the formalism used to describe the dynamics holds for any arbitrary Gaussian phase randomization process our conclusions hold in general for these types of models \cite{Paper}. 

For $\Delta t\!>\!\bar{t}$ and for $s\!>\!2$, non-Markovian effects become relevant in the overall free dynamics and the amount of coherence preservation will depend on a combined effect of both the presence of information back-flow connected to the unperturbed dynamics and effect of the pulses. A strong dependence on the time interval $\Delta t$ as well as on $t_f$ emerges in this case from numerical studies. This can be traced back to the fact that the unperturbed dynamics for $t$ is characterized by subsequent time intervals in which the information flow changes sign. Hence pulses will enhance decoherence or preserve coherence depending on whether they occur in a time interval in which information flow is positive or negative, respectively.

\section{Non-Markovianity engineering by dynamical decoupling}

Markovian open quantum systems have been extensively studied and are very well characterized. For Markovian dynamics fulfilling the semigroup property \cite{bp}, the Lindblad-Gorini-Kossakowski-Sudarshan theorem identifies the general form of master equation leading to a physical evolution of the system. The Monte Carlo wave function approach provides both a powerful numerical technique to study the dynamics of Markovian systems and a deep interpretation in terms of quantum jumps for individual quantum systems, like ions or cavity modes. Quantum state diffusion methods allow to unravel the dynamics in terms of homodyne or heterodyne measurements on the environment.
 \begin{figure}[h]
\begin{centering}
\includegraphics[width=0.85\textwidth]{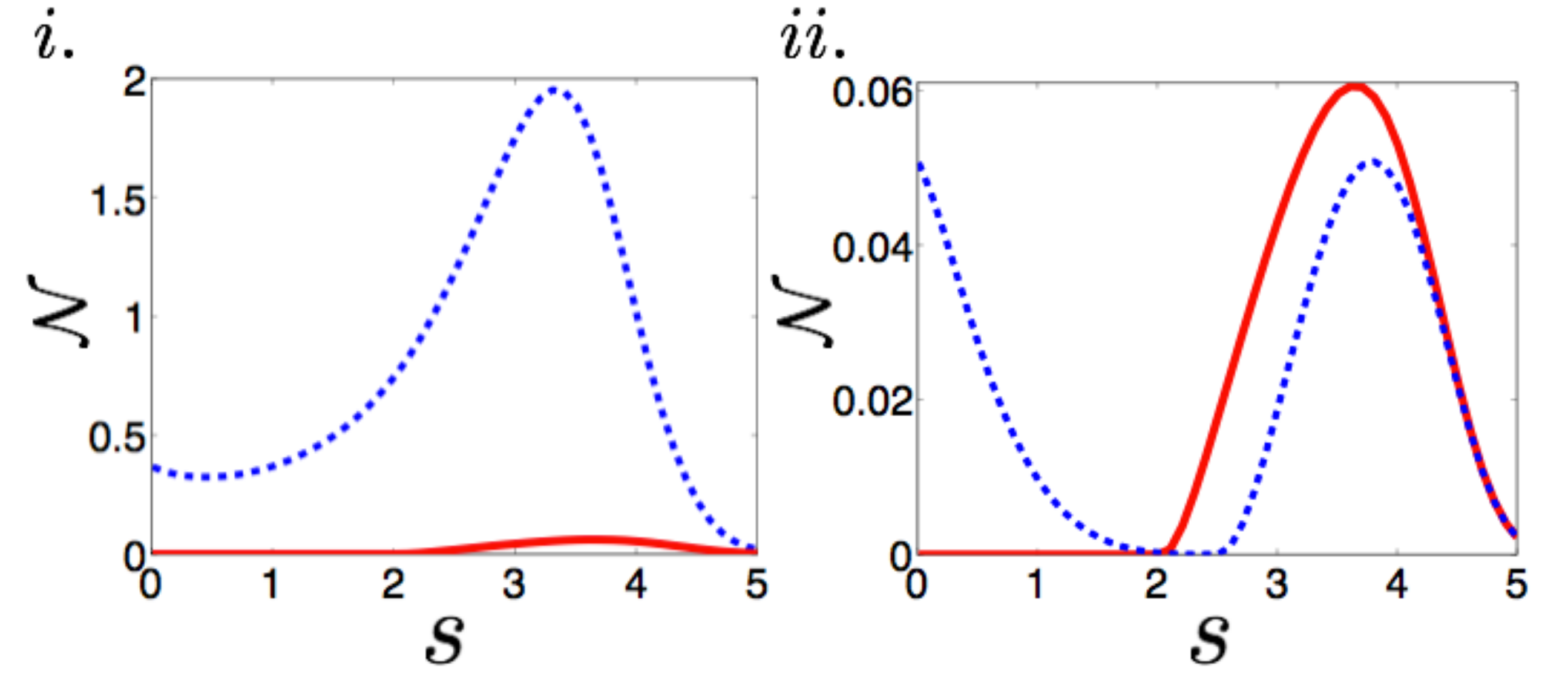}
\caption{(Color online) Non-Markovianity measure against the Ohmicity parameter $s$ for the dynamically decoupled system (blue dotted line) for time intervals i) $\Delta t=0.3$ and ii. $\Delta t=3$ (in units of $\omega_c^{-1}$) and the non-Markovianity for the free system (red solid line). The final pulse is applied at time $\omega_c t_f=9$.  We draw attention to the fact that $\mathcal{N}\neq 0$ for all values of $s$ when the system is subject to DD but may take very small values as shown in ii) for $2\leq s<2.6$.  }
\label{figure01}
\end{centering}
\end{figure} 

For non-Markovian open quantum systems many fundamental questions are still open. The generalisation of the Lindblad-Gorini-Kossakowski-Sudarshan theorem to even simple non-Markovian systems is still an open problem. The existence of a measurement scheme interpretation guaranteeing a physical meaning to individual trajectories is still under investigation. The extension of the Monte Carlo wave function approach is only known for certain classes of time-local master equations \cite{NMQJ}. The first experimental studies aimed at characterizing non-Markovian dynamics have only recently been conducted \cite{exp1}-\cite{exp4}. This witnesses the interest in developing techniques for engineering non-Markovian dynamics to be used as testbeds for experimental and theoretical investigations.

The results of section 4 show that, in addition to its traditional employment as a method to hamper decoherence, DD can be exploited as a simple tool for engineering non-Markovian dynamics.
A Markovian open system will, indeed, always become non-Markovian when subject to PDD. More in general, PDD will change the non-Markovian character of the open system, whether its free dynamics was Markovian or not. Yet, the details of the pulse-induced non-Markovianity will depend on both the pulsing parameters (e.g., the pulse spacing) and the environmental parameters (e.g., the Ohmicity parameter). 

In this section, we investigate the non-Markovianity induced by PDD by comparing the non-Markovianity measure $\mathcal{N}$ in absence and presence of pulses, in both the short-pulse and the long-pulse regimes introduced in section 3. As we noticed there, in the short-pulse regime, for any value of the Ohmicity parameter $s$, the effect of the pulses is to create non-Markovianity by inducing information back-flow, when it was initially absent, or in any case to increase the non-Markovian character. This can be seen in Figure 4 i. In the long-pulsing regime the situation is more variegated as pulses can also, under certain conditions, decrease the non-Markovian character of the dynamics, as shown in the example of figure 4 ii. 

To conclude, any system subject to PDD will provide a testbed for further investigating non-Markovian dynamics by inducing information back-flow, independently of the value of the Ohmicity parameter $s$. It is worth noticing once again that engineering non-Markovianity here refers specifically to the information-theoretical approach which has been proven useful for quantum technologies \cite{QKD, P6, QT1, QT2}. In this sense our results should not simply understood as another variant of the well known idea that DD modifies the reservoir spectrum by making it more structured. On the contrary, they are an exploration on the ability to controllably modify and enhance quantities such as the channel capacities, mutual information, coherent information, Fisher information, etc. This in turn provides a way to control the efficiency of quantum communication protocols, quantum metrology, and work extraction, just to mention a few.

\section{Long-Time Dynamics}
While in the previous sections we have shed light on the connection between the spectral density shape  and the dynamical decoupling effectiveness for short times, we now turn our focus to the {\it asymptotic} behavior of the pulsed system. With reservoir engineering in mind, we investigate the connection between the stationary coherences and the form of the spectral density function (specifically, the value of the Ohmicity parameter $s$). More precisely, we study which value of $s$ yields maximum long-time stationary coherences, for given pulses time spacing $\Delta t$ and number of pulses $n$. 
 We consider specifically the case in which the DD sequence stops at a finite $t_f$, after which the system is subjected to the usual decoherence arising from its unavoidable environment. The case in which $t_f$ goes to infinity is neither analytically nor numerically treatable.

We begin by noticing that in absence of pulses the phenomenon of coherence trapping occurs for $s > 1$, while for $s \le 1$ coherences are asymptotically lost as $t \rightarrow \infty$. The controlled coherence function for long times $\Gamma_N(\infty)$, recalling $\Gamma(t)=\Gamma_N(t)$ for $t_f<t$, where $N$ is the total number of pulses, is as follows (using $\Gamma_0(\infty)=\Gamma[s-1]$ \cite{Greg}):
\eqa
\Gamma_N(\infty)&=&2\sum_{m=1}^N(-1)^{m+1}\Gamma_0(t_m)\nn\\&+&4\sum_{m=2}^N\sum_{j<m}\Gamma_0(t_m-t_j)(-1)^{m-1+j}\nn\\&+&2\sum_{m=1}^N(-1)^{m+N}\Gamma[s-1]+(-1)^N\Gamma[s-1].\nn\\
\eeqa
Figure 5 shows how the maximum stationary coherence in the DD scheme depends on the pulsing interval as a function of $s$ and for different numbers of pulses $n$. From \cite{Greg} we know that, for the unperturbed system, the optimal Ohmicity parameter, i.e., the value of $s$ leading to maximum long-time coherences, lies in the non-Markovian parameter range ($s\simeq\!2.46$). In presence of pulses, this continues to be true, independently of the number of pulses $n$, only in the short-pulsing regime, as one can see for the exemplary value $\omega_c \Delta t = 0.3$ (first column of dots in figure 5). 

 It is interesting to compare the stationary coherences to the coherences present at time $t_f$, i.e., at the end of the pulse sequence. We have noted in section 4 that, for finite $t_f$, Markovian reservoirs lead to better coherence preservation when the interval between pulses is short.  Hence the choice of optimal $s$ depends on whether we are interested in the coherences at the end of the pulse sequence or in the asymptotic stationary coherences.

As shown in figure 5, when the interval between pulses increases, Markovian reservoirs become better suited to long-time coherences preservation for most values of $n$, with the only exception of the somewhat special case $n=1$ (red dots) and also $n=2$ (orange dots) for $\omega_c\Delta t=3$.

\begin{figure}[h]
\begin{centering}
\includegraphics[width=0.65\textwidth]{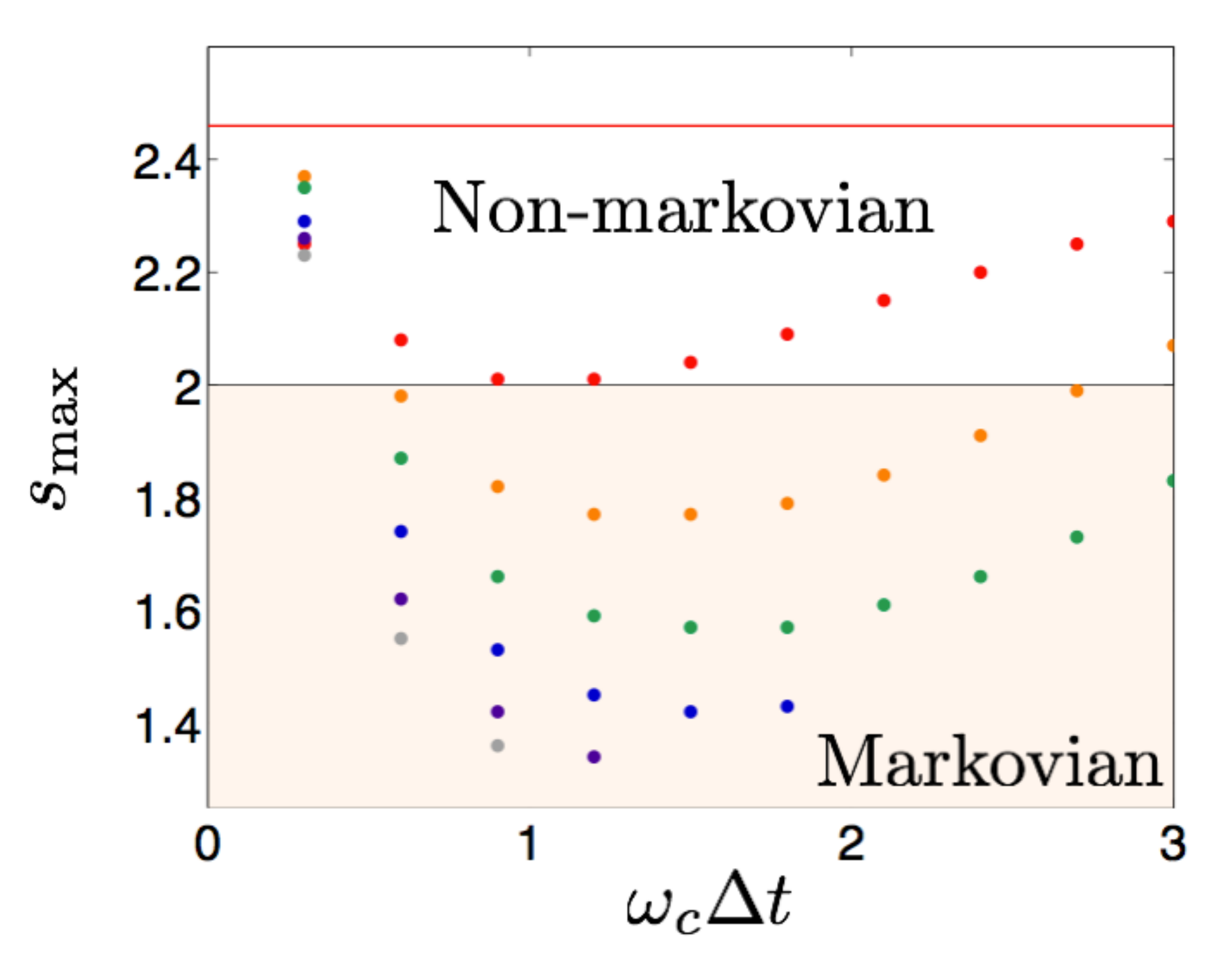}
\caption{(Color online) Ohmic parameter $s$ required to achieve maximum stationary coherence as a function of the (rescaled) pulse interval spacing $\Delta t$ for $n=1$ (red dots), $n=2$ (orange), $n=5$, (green), $n=10$ (blue), $n=20$ (purple) and $n=30$ (grey). The free Markovian (non-Markovian regime) corresponds to the shaded (unshaded) region, while the red line shows the value of the Ohmic parameter for the free stationary coherence. We have ignored stationary coherences with maximum values below the order of $10^{-4}$.} 
\label{s}
\end{centering}
\end{figure} 

\section{Conclusions}
To conclude, our results provide indications on how one should engineer an environment which is optimal for dynamical decoupling techniques. We have explored the connection between information flow and dynamical decoupling to shed light on the phenomena responsible for revivals in the coherence. Having efficient error correction in mind, we have paid special attention to the short-pulses regime. In this case, we have found that a Markovian environment is necessary to optimize the DD performance. However, the highest preserved coherence at the end of the decoupling sequence is not necessarily the highest stationary coherence (long time limit), i.e., the optimal Ohmicity parameter is not the same for the two regimes. Our work provides the first exploration of the interplay between non-Markovianity in terms of information flow, as defined in \cite{BLP, P6, MI, RHP}  and the efficiency of DD schemes. With a shift in perspective, it also indicates how dynamical decoupling techniques can be harnessed to engineer quantum non-Markovianity and control it.

\section*{Acknowledgments}{We acknowledge financial support from the Horizon 2020 EU collaborative project QuProCS (Grant Agreement 641277). }

\end{document}